\documentclass[twocolumn,showpacs,amsmath,amssymb,nofootinbib,aps,prd,longbibliography,10pt]{revtex4-1}
\usepackage{dcolumn}
\usepackage{bm}
\usepackage{graphicx}
\usepackage[dvipsnames]{xcolor}

\usepackage{amsthm}
\usepackage[normalem]{ulem}

\newcommand{\be}{\begin{equation}}
\newcommand{\ee}{\end{equation}}
\newcommand{\ba}{\begin{eqnarray}}
\newcommand{\ea}{\end{eqnarray}}
\newcommand{\non}{\nonumber}
\newcommand{\n}[1]{\label{#1}}
\newcommand{\eq}[1]{(\ref{#1})}
\newcommand{\hh}{\, ,\hspace{0.25cm}}
\newcommand{\hhh}{\, ,\hspace{0.5cm}}
\newcommand{\BM}[1]{{\mbox{\boldmath $#1$}}}

\newcommand{\h}[1]{\hat{#1}}

\begin{document}

\title{Gravitational Faraday and spin-Hall effects of light: Local description}
\author{Andrey A. Shoom}
\email{andrey.shoom@aei.mpg.de}
\affiliation{Max Planck Institute for Gravitational Physics (Albert Einstein Institute),
Leibniz Universit\"at Hannover, Callinstr. 38, 30167, Hannover, Germany}

\date{\today}

\begin{abstract} 
 
A gravitational field can cause a rotation of the polarisation plane of light. This phenomenon is known as the gravitational Faraday effect. It arises due to different spin-orbit interaction of left- and right-handed circularly polarised components of light. Such an interaction also causes transverse displacement in the light trajectory, in opposite directions for each component. This phenomenon is known as the gravitational spin-Hall effect of light. We study these effects in a local inertial frame in arbitrary vacuum space-time and show that they are observer dependent and arise due to interaction of light polarisation with a local gravitomagnetic field measured by observer. Thus, to address the effects to a gravitational field alone, one has to consider zero angular momentum observers. 

\end{abstract}

\maketitle

\section{Introduction}

A study of light propagation in strong gravitational fields in the geometric optics approximation opens up very interesting observational possibilities, such as light deflection and shadows of gravitating objects \cite{Perlick:2021aok,Bisnovatyi-Kogan:2022ujt,Tsupko:2022kwi,Aratore:2024bro}. Going beyond the geometric optics formalism should be more informative and interesting. In this work we shall study the gravitational Faraday and spin-Hall effects of light in arbitrary gravitational field (vacuum space-time) in a local inertial frame. These effects are related to each other and arise due to interaction of light polarisation (helicity) and a gravitomagnetic field defined in the observer's frame.

The gravitational Faraday effect is a rotation of the polarisation plane of an electromagnetic wave (light) propagating in a gravitational field, for example, near a rotating black hole. A study and observation of this effect have quite a long history \cite{Skrotskii,Plebanski,God,GF2,GF3,GF4,GF5,GF6,CariniRuffini,Perlick_1993,GF7,GF8,GF9,Halilsoy:2006ev,GF10,Frolov:2011mh,Ghosh}. In these works a stationary gravitational field of a rotating (Kerr) black hole was considered. A stationary space-time possesses time-like Killing vector field, which can be adapted to define observers who share the same space-like hypersurface, that allows to consider a space-time decomposition globally and describe the gravitational Faraday effect. In case of a general space-time there is no preferred global time-like vector field which could be used to define observers. To overcome this problem, here we take a local approach and consider an inertial observer who can perform experiments in a local laboratory having a fixed spatial coordinate location of all the laboratory equipment bolted into the laboratory frame. The laboratory should be sufficiently large to capture the space-time curvature and sufficiently small to define a local regular coordinate grid. The local coordinates are the Fermi normal coordinates, which naturally serve this purpose.           

There is a related gravitational spin-Hall effect resulting in transverse polarisation-dependent displacement of light trajectory due to the interaction between its polarisation and a gravitomagnetic field. As a result, a linearly polarised light splits into left- and right-handed circularly polarised components propagating along different paths. Originally, this effect was studied in the weak field approximation for light propagating in a stationary gravitational field near a rotating gravitational body \cite{Mashhoon:1973,Mashhoon:1974,Mashhoon:1974a,Mashhoon:1975ki}. It was shown that left- and right-handed circularly polarised light components get scattered differently. To describe this effect in a strong stationary gravitational field, the so-called modified geometric optics formalism was introduced \cite{Frolov:2011mh}. Then, the formalism was applied to describe scattering of polarised light propagating in the stationary space-time of a Kerr black hole \cite{Frolov:2012zn}. Gravitational spin-Hall effect of light was studied in later works without its relation to the gravitational Faraday effect \cite{Yoo,Dolan:2018,Oancea:2020khc,Frolov:2020uhn}. Here we shall consider a local description of the gravitational Faraday and spin-Hall effects in the frame defined by the Fermi normal coordinates. In Sec.~II we review the basic laws of geometric optics. In Sec.~III an inertial observer's frame is defined via the Fermi normal coordinates. Section~IV contains a study of the gravitational Faraday effect in the inertial observer's frame. In Sec.~V we study the related gravitational spin-Hall effect of light. In Sec.~VI we discuss a role of observer in detection of these effects. Section VII contains summary and discussion of derived results.
 
Here we use a system of units in which $G=c=1$ and conventions adopted in the book \cite{MTW}.  

\section{Geometric Optics}

Electromagnetic waves propagating in a gravitational field are described by the Maxwell equations in a curved space-time background. For waves that are highly monochromatic over some space-time regions, an asymptotic short-wave (geometric optics) approximation is often used \cite{MTW}. The approximation captures basic characteristics of light propagation in a curved space-time background defined by metric $g_{\alpha\beta}$. In what follows, we shall consider a vacuum space-time and consider the source-free Maxwell wave equation for the vector potential $A^{\alpha}$ in the Lorenz gauge
\be\n{2.1}
A^{\alpha}_{\,\,\,;\alpha}=0\,,
\ee
which reads
\be\n{2.2}
A^{\alpha;\beta}_{\,\,\,\,\,\,\,\,;\beta}=0\,.
\ee
Here and in what follows, the semicolon stands for the covariant derivative associated with the used space-time metric. 

The geometric optics approximation begins with a splitting of the vector potential into a rapidly changing real phase, called the {\em eikonal} $\vartheta$, and a slowly changing complex amplitude,
\be\n{2.3}
A^{\alpha}=\Re\{(a^{\alpha}+\varepsilon b^{\alpha}+...)e^{i\vartheta/\varepsilon}\}\,,
\ee
where $\varepsilon\ll1$ is a dummy expansion parameter used to track order of terms: a term with $\varepsilon^n$, for some integer $n$, varies as $(\lambdabar/{\cal L})^n$, where $\lambdabar/{\cal L}\ll1$. Here $\lambdabar$ is the reduced wavelength (wavelength/$2\pi$) and ${\cal L}$ is the minimal of the two characteristic scales\textemdash the curvature radius of the wave front, or the length of a wave packet, and the local curvature radius of the space-time. 

At this stage, it is convenient to introduce quantities in terms of which sought laws of the geometric optics are formulated. They are the following. The eikonal gradient and metrically related to it the wave vector,
\be\n{2.4}
k_{\alpha}\equiv\vartheta_{,\alpha}\hhh k^{\alpha}=g^{\alpha\beta}k_{\beta}=dx^{\alpha}/d\lambda\,,
\ee
tangent to the light ray $\Gamma\!\!:x^{\alpha}=x^{\alpha}(\lambda)$, where $\lambda$ is affine parameter of the ray. The scalar amplitude, 
\be\n{2.5}
a\equiv(a^{\alpha}\bar{a}_{\alpha})^{1/2}\,,
\ee
and the unit complex polarisation vector, 
\be\n{2.6}
f^{\alpha}\equiv a^{\alpha}/a\hh f^{\alpha}f_{\alpha}=0\hh f^{\alpha}\bar{f}_{\alpha}=1\,.
\ee
Here and in what follows, the bar stands for complex conjugation. Substituting the vector potential \eq{2.3} into the Lorenz gauge condition \eq{2.1} and the wave equation \eq{2.2} and collecting the leading terms of order $\varepsilon^{-2}$ and $\varepsilon^{-1}$ we derive the fundamental laws of geometric optics, 
\ba
k^{\alpha}k_{\alpha}=0&\hh&k^{\beta}k^{\alpha}_{\,\,\,;\beta}=0\,,\n{2.7}\\
k^{\alpha}f_{\alpha}=0&\hh&k^{\beta}f^{\alpha}_{\,\,\,;\beta}=0\,,\n{2.8}\\
&&\hspace{-1cm}(a^{2}k^{\alpha})_{;\alpha}=0\,.\n{2.9}
\ea
These laws imply that light rays are space-time null geodesics \eq{2.7}, the polarisation vector is orthogonal to the light ray and parallel-propagated along it \eq{2.8}, and the vector $a^2k^{\alpha}$ is a conserved current, which defines the adiabatically conserved number of light rays, or in quantum language, the number of photons \eq{2.9}. Propagation of light in a curved space-time is described by the laws of geometric optics approximately. Note that in this description polarisation of light does not affect its path. Thus, in order to describe the gravitational spin-Hall effect of light, one has to modify the standard laws of geometric optics, which is done in Sec.~V.  
 
\section{Inertial observer}

Consider an {\em inertial} observer moving along time-like geodesic $O(\tau)$ with 4-velocity $u^{\alpha}(\tau)$ in a gravitational field $g_{\alpha\beta}$. Fermi normal coordinates \cite{Manasse:1963zz},
\be\n{3.1}
x^{\h\alpha}=(x^{\h0}=\tau\,,\,x^{\h i}\,; i=1,2,3)\,
\ee
associated with this observer, are used to describe local experiments performed in the observer frame. The spatial coordinates $x^{\h i}$ define the proper distance $\ell$ from some event on $O(\tau)$ to a spatially separated event in its vicinity via $x^{\h i}=\ell\,v^{\h i}$, where $v^{\h\alpha}$ is a space-like unit vector defined at $O(\tau)$ and orthogonal to $u^{\alpha}$. The space-time metric in the vicinity of $O(\tau)$ computed up to third order in $\ell$ reads \cite{LiNi,DolKhri},
\ba\n{3.2}
ds^2&=&-\left(1+R_{\h0\h i\h0\h j}x^{\h i}x^{\h j}+\tfrac{1}{3}R_{\h0\h i\h0\h j;\h k}x^{\h i}x^{\h j}x^{\h k}\right)(dx^{\h0})^2\non\\
&-&2\left(\tfrac{2}{3}R_{\h0\h k\h i\h j}x^{\h k}x^{\h j}+\tfrac{1}{4}R_{\h0\h k\h i\h j;\h l}x^{\h k}x^{\h j}x^{\h l}\right)dx^{\h0}dx^{\h i}\non\\
&+&\left(\delta_{\h i\h j}-\tfrac{1}{3}R_{\h i\h k\h j\h l}x^{\h k}x^{\h l}-\tfrac{1}{6}R_{\h i\h k\h j\h l;\h m}x^{\h k}x^{\h l}x^{\h m}\right)dx^{\h i}dx^{\h j}\non\\
&+&{\cal O}(\ell^4)dx^{\h\alpha}dx^{\h\beta}\,,
\ea
where $R_{\h\alpha\h\beta\h\gamma\h\delta}$ and $R_{\h\alpha\h\beta\h\gamma\h\delta;\h\mu}$ are evaluated on $O(\tau)$, that is they depend on the observer's proper time $\tau$ only. This metric approximation is valid if the proper distance $\ell$ is such that
\be\n{3.3}
\ell\ll\text{min}\left\{\frac{1}{|R_{\h\alpha\h\beta\h\gamma\h\delta}|^{1/2}}\hh\frac{|R_{\h\alpha\h\beta\h\gamma\h\delta}|}{|R_{\h\alpha\h\beta\h\gamma\h\delta,\h\sigma}|}\right\}\,.
\ee
This condition implies that the spatial geodesics in the vicinity of $O(\tau)$ do not intersect.\footnote{One may consider higher order terms in the metric expansion, which can be constructed according to the procedure described in \cite{LiNi,DolKhri}.}

To define the observer's local laboratory we consider a local non-geodesic congruence of time-like world lines $x^{\h i}=\text{const}$ associated with the observer. The congruence defines world lines of each spatial point indicating location of an apparatus in the laboratory. Using the metric \eq{3.2} we derive 4-velocity vectors tangent to the world lines in the corresponding order of the used approximation,
\be\n{3.4}
u^{\h\alpha}(x^{\h\beta})=\left(1-\tfrac{1}{2}R_{\h0\h i\h0\h j}x^{\h i}x^{\h j}-\tfrac{1}{6}R_{\h0\h i\h0\h j;\h k}x^{\h i}x^{\h j}x^{\h k}\right)\delta^{\h\alpha}_{\h0}\,,
%\\
%u_{\h\alpha}(x^{\h\beta})&\approx&-\left(1+\frac{1}{2}R_{\h0\h i\h0\h j}x^{\h i}x^{\h j}\right)\delta^{\h0}_{\h\alpha}-\frac{2}{3}R_{\h0\h k\h i\h j}x^{\h k}x^{\h j}\delta^{\h i}_{\h\alpha}\,,\non
\ee
Along every world line we define an orthonormal local tetrad
\ba\n{3.5}
&&\{e^{\h\alpha}_{\,\,\,(0)}=u^{\h\alpha}(x^{\h\beta})\,,\, e^{\h\alpha}_{\,\,\,(a)}\,; a = 1,2,3\}\,,\\
&&\eta_{(a)(b)}=e^{\h\alpha}_{\,\,\, (a)}e^{\h\beta}_{\,\,\, (b)}\,g_{\h\alpha\h\beta}=\text{diag}(-1,1,1,1)\,.
\ea
which we shall use to define measurements at a given point in the laboratory frame.

\section{Gravitational Faraday effect}

In this section we study the gravitational Faraday effect. We begin with writing down the complex polarisation vector in the laboratory frame,
\be\n{4.1}
f^{\h\alpha}=e^{-i\varphi_{\sigma}}m^{\h\alpha}\,,
\ee
where $m^{\h\alpha}$ is a unit complex vector defined in the local tetrad \eq{3.5} as follows:
\be\n{4.2}
m^{\h\alpha}\equiv\tfrac{1}{\sqrt{2}}(e^{\h\alpha}_{\,\,\, (1)}+i\sigma e^{\h\alpha}_{\,\,\, (2)})\hh \bar{m}^{\h\alpha}\equiv\tfrac{1}{\sqrt{2}}(e^{\h\alpha}_{\,\,\, (1)}-i\sigma e^{\h\alpha}_{\,\,\, (2)})\,,
\ee
where the parameter $\sigma=\pm1$ specifies polarisation, with $`+$' for the right- and `$-$' for the left-handed circularly polarised light \cite{MTW}. Note that the polarisation vector $f^{\h\alpha}$ is defined up to an additive of the wave vector $k^{\h\alpha}$. We fix this gauge freedom by imposing that the complex vector $m^{\h\alpha}$ is orthogonal to the local spatial direction of the null ray, defined by the tetrad vector $e^{\h\alpha}_{\,\,\, (3)}$. For the defined complex vectors the following relations hold:
\be\n{4.3}
m^{\h\alpha}m_{\h\alpha}=0\hhh m^{\h\alpha}\bar{m}_{\h\alpha}=1\hhh m^{\h\alpha}k_{\h\alpha}=0\,.
\ee 
{\em Polarisation phase} $\varphi_{\sigma}$ defines a change due to the gravitational field in the natural rotation of the polarisation vector along a null ray with respect to the complex basis $\{m^{\h\alpha}, \bar{m}^{\h\alpha}\}$. Defined this way polarisation phase corresponds to positive values of $\varphi_{0}$ (see the expression \eq{4.12} below) measured in the anti-clockwise direction from $e^{\h\alpha}_{\,\,\, (1)}$. There is also gauge freedom due to rotations in the 2-dimensional plane perpendicular to the local spatial direction of the null ray. According to \eq{4.1}, it implies that the polarisation phase is defined up to an arbitrary angle of rotation of the basis $\{m^{\h\alpha}, \bar{m}^{\h\alpha}\}$. We shall fix this gauge later, when we define a propagation law for the basis along the null ray.  

Applying the propagation equation for the polarisation vector  \eq{2.8} to the expression \eq{4.1}, we derive an evolution equation for the polarisation phase along the null ray,
\be\n{4.4}
k^{\h\alpha}\varphi_{\sigma,\h\alpha}=-i\bar{m}_{\h\alpha}k^{\h\beta}m^{\h\alpha}_{\,\,\,;\h\beta}\,.
\ee
The next step is to define a propagation law for $m^{\h\alpha}$ along the null ray. This is done by a local decomposition of the space-time into space and time. By the construction, there is a local tetrad \eq{3.5} defined at every point of a null ray. For a given $e^{\h\alpha}_{\,\,\,(0)}=u^{\h\alpha}$ there is a local 3-dimensional subspace $\Sigma_{u}$ of a tangent space orthogonal to $u^{\h\alpha}$. A vector from the tangent space can be projected into the subspace $\Sigma_{u}$ by means of the projection operator
\be\n{4.5}
\Pi_{\h\beta}^{\h\alpha}=\delta_{\h\beta}^{\h\alpha}+u^{\h\alpha}u_{\h\beta}\,.
\ee
Applying the projection operator to $k^{\h\alpha}$ we construct the unit space-like vector $n^{\h\alpha}=e^{\h\alpha}_{\,\,\, (3)}$ that defines the spatial direction of a light ray, as we mentioned above. Accordingly, we have
\be\n{4.6}
k^{\h\alpha}=\omega(u^{\h\alpha}+n^{\h\alpha})\,,
\ee
where $\omega=-k_{\h\alpha}u^{\h\alpha}$ is the local angular frequency of light. The decomposition allows to express a propagation of $m^{\h\alpha}$ along $k^{\h\alpha}$ by defining its propagation along $u^{\h\alpha}$ and $n^{\h\alpha}$, 
\be\n{4.7}
k^{\h\beta}m^{\h\alpha}_{\,\,\,;\h\beta}=\omega(u^{\h\beta}m^{\h\alpha}_{\,\,\,;\h\beta}+n^{\h\beta}m^{\h\alpha}_{\,\,\,;\h\beta})\,.
\ee
To measure unambiguously the polarisation phase evolution along the light ray \eq{4.4}, we require firstly that $m^{\h\alpha}$ does not rotate about the null ray spatial trajectory defined by its tangent vector $n^{\h\alpha}$. If the trajectory were a space-like geodesic, then to fulfil this condition, $m^{\h\alpha}$ had to be parallel transported along it. In a general case, it has to undergo a Fermi transport, which implies that the term $n^{\h\beta}m^{\h\alpha}_{\,\,\,;\h\beta}$ lies in a 2-dimensional plane spanned by $n^{\h\alpha}$ and the acceleration $n^{\h\beta}n^{\h\alpha}_{\,\,\,;\h\beta}$. But because initially $m^{\h\alpha}$ is orthogonal to $n^{\h\alpha}$, the term $n^{\h\beta}m^{\h\alpha}_{\,\,\,;\h\beta}$ is collinear to $n^{\h\alpha}$. Therefore, then we plug \eq{4.7} into \eq{4.4}, this term gives zero contribution to the polarisation phase evolution. Second, we require that $m^{\h\alpha}$ preserves its original orientation when transported along a world line defined by $u^{\h\alpha}$. This means that $m^{\h\alpha}$ has to corotate with the laboratory congruence (see \S 13.6 in \cite{MTW}), that is
\be\n{4.8}
u^{\h\beta}m^{\h\alpha}_{\,\,\,;\h\beta}=(u^{\h\alpha}w_{\h\beta}-w^{\h\alpha}u_{\h\beta})m^{\h\beta}-\varepsilon_{\h\alpha\h\beta\h\gamma\h\delta}u^{\h\beta}\Omega^{\h\gamma}m^{\h\delta}\,.
\ee 
Here $w^{\h\alpha}=u^{\h\beta}u^{\h\alpha}_{\,\,\,;\h\beta}$ is the 4-acceleration of the local laboratory equipment and
\be\n{4.9}
\Omega^{\h\alpha}=\tfrac{1}{2}\varepsilon^{\h\alpha\h\beta\h\gamma\h\delta}u_{\h\beta}u_{\h\gamma;\h\delta}\,,
\ee
is its angular velocity (vorticity), which arise due to a fixed spatial distance and orientation with respect to the inertial observer. The Levi-Civita (pseudo) tensor $\varepsilon_{\h\alpha\h\beta\h\gamma\h\delta}$ is normalised as $\varepsilon_{(0)(1)(2)(3)}=+1$. The propagation law \eq{4.8} fixes the rotational gauge freedom discussed earlier. Plugging \eq{4.7} and \eq{4.8} into \eq{4.4} and using properties of the Levi-Civita tensor we derive the polarisation phase evolution,
\be\n{4.10}
k^{\h\alpha}\varphi_{\sigma,\h\alpha}=-\sigma\Omega_{\h\alpha}k^{\h\alpha}\,,
\ee
which allows us to compute evolution of the polarisation phase along the null ray $\Gamma\!\!:x^{\alpha}=x^{\alpha}(\lambda)$,
\be\n{4.11}
\varphi_{\sigma}=-\sigma\int_{\Gamma}\Omega_{\h\alpha}k^{\h\alpha}d\lambda=-\sigma\int_{\Gamma}
\Omega_{\h\alpha}dx^{\h\alpha}\,.
\ee 
Finally, we consider a linearly-polarised light as a superposition of its left- and right-handed circularly polarised components. The angle of rotation of the linear polarisation vector $f^{\h\alpha}_{0}=(f^{\h\alpha}+\bar{f}^{\h\alpha})/\sqrt{2}$ measured along the light ray ${\Gamma}$ is 
\be\n{4.12}
\varphi_{0}=-\int_{\Gamma}\Omega_{\h\alpha}k^{\h\alpha}d\lambda\,.
\ee
This rotation is known as the {\em gravitational Faraday effect of light.}

To derive an expression for the gravitational Faraday effect measured in the laboratory frame \eq{3.2} we compute vorticity \eq{4.9} of the congruence \eq{3.4} within the given order of approximation,
\be\n{4.13}
\Omega^{\h\alpha}=-\tfrac{1}{4}\epsilon^{\h 0\h i\h j\h k}\left(2R_{\h 0\h l\h j\h k}x^{\h l}+R_{\h 0\h l\h j\h k;\h m}x^{\h l}x^{\h m}\right)\delta^{\h\alpha}_{\h i}\,,
\ee
where $\epsilon^{\h 0\h 1\h 2\h 3}=-1$ is the Levi-Civita symbol. In a vacuum space-time the Riemann tensor coincides with the Weyl tensor, that allows us to express the vorticity in the following way: 
\be\n{4.14}
\Omega_{\h\alpha}=-\left({\cal B}_{\h i\h j}x^{\h j}+\tfrac{1}{2}{\cal B}_{\h i\h j;\h k}x^{\h j}x^{\h k}\right)\delta^{\h i}_{\h\alpha}\,.
\ee 
Here the gravitomagnetic field ${\cal B}_{\h i\h j}(\tau)$ and its gradient ${\cal B}_{\h i\h j;\h k}(\tau)$ are measured along the observer's world line. The gravitomagnetic field is defined as follows \cite{PhysRevD.33.915,Membrane,GRSolutions}\footnote{Note that \cite{GRSolutions} uses different sign convention for the gravitomagnetic field.}:
\be\n{4.15}
{\cal B}_{\h\alpha\h\beta}\equiv^{\,\,\,*\!\!\!}C_{\h\gamma\h\alpha\h\beta\h\delta}u^{\h\gamma}u^{\h\delta}\hh^{*\!}C_{\h\alpha\h\beta\h\gamma\h\delta}=\frac{1}{2}\varepsilon_{\h\alpha\h\beta\h\mu\h\nu}C^{\h\mu\h\nu}_{\,\,\,\,\,\,\,\h\gamma\h\delta}\,,
\ee
such that
\be\n{4.16}
{\cal B}_{\h\alpha\h\beta}={\cal B}_{\h i\h j}\delta^{\h i}_{\h\alpha}\delta^{\h j}_{\h\beta}\hhh {\cal B}_{\h i\h j}={\cal B}_{\h j\h i}\hhh {\cal B}^{\h i}_{\,\,\,\h i}=0\,,
\ee
We define its gradient as
\be\n{4.17}
{\cal B}_{\h\alpha\h\beta;\h\mu}\equiv^{\,\,\,*\!\!\!}C_{\h\gamma\h\alpha\h\beta\h\delta;\h\mu}u^{\h\gamma}u^{\h\delta}\,.
\ee
Then the expression for the gravitational Faraday effect \eq{4.12} takes the following form:
\be\n{4.18}
\varphi_{0}=\int_{\Gamma}\left({\cal B}_{\h i\h j}x^{\h j}+\tfrac{1}{2}{\cal B}_{\h i\h j;\h k}x^{\h j}x^{\h k}\right)k^{\h i}d\lambda\,.
\ee
To compute the polarisation phase one has to find null geodesics in the frame \eq{3.2} within the order of approximation \eq{3.3}. 

According to the approximation condition \eq{3.3}, the gravitational Faraday rotation $\varphi_{0}\ll1$ for a light ray traveling across the local laboratory. However, as it is seen from \eq{4.18}, changing spatial direction of a light ray to the opposite and assuming that vorticity \eq{4.14} does not change its sign during the observation time $\tau$, the rotation of the polarisation plane changes from clockwise to the counterclockwise (or vice versa), as seen from the light ray direction, thereby keeping the same direction of rotation in the laboratory frame. Thus, by setting up a set of mirrors and making the light to go back and forth across the laboratory allows for an accumulation of the angle of rotation $\varphi_{0}$, that after sufficiently large time $\tau$ may result in a large net value of $\varphi_{0}$. 

\section{Gravitational spin-Hall effect}

The consequence of the gravitational Faraday effect is so that due to a coupling between light polarisation (or helicity) and angular momentum of its trajectory, the right- and left-polarised light components propagate along different paths. This phenomenon is called the gravitational spin-Hall effect of light. 

In order to describe this effect we have to go beyond the standard eikonal approach \eq{2.3}. The reason is that the standard eikonal expansion is not uniformly valid in its domain, because for short distances a contribution of polarisation evolution to propagation of light is of order $\varepsilon$, but the polarisation effects accumulate along light ray trajectory and for sufficiently large distances of order $\varepsilon^{-1}$ they become of order $\varepsilon^0$. Thus, for an eikonal expansion to be uniformly valid everywhere, we have to take into account polarisation evolution. Namely, we define the {\em combined eikonal},
\be\n{5.1}
\tilde{\vartheta}\equiv\vartheta+\varepsilon\varphi_{\sigma}\,.
\ee
Then following the same steps taken in deriving the laws of geometric optics we find that the gradient of the modified eikonal is
\be\n{5.2}
\tilde{k}_{\h\alpha}\equiv\tilde\vartheta_{,\h\alpha}=\vartheta_{,\h\alpha}+\varepsilon\varphi_{\sigma,\h\alpha}\,,
\ee
and the modified wave vector $\tilde{k}^{\h\alpha}$ is null. It implies that $\tilde{k}^{\h\beta}\tilde{k}_{\h\beta;\h\alpha}=0$. From the expressions \eq{4.11}, \eq{5.2} and the gradient equality $\vartheta_{;\h\beta\h\alpha}=\vartheta_{;\h\alpha\h\beta}$ it follows that
\be\n{5.3}
\tilde{k}_{\h\beta;\h\alpha}=\tilde{k}_{\h\alpha;\h\beta}-\varepsilon\sigma\Phi_{\h\alpha\h\beta}\,,
\ee
where we defined
\be\n{5.4}
\Phi_{\h\alpha\h\beta}\equiv\Omega_{\h\beta;\h\alpha}-\Omega_{\h\alpha;\h\beta}\,.
\ee
As a result, we derive the following equation for null rays:
\be\n{5.5}
\tilde{k}^{\h\beta}\tilde{k}_{\h\alpha;\h\beta}=\varepsilon\sigma\Phi_{\h\alpha\h\beta}\tilde{k}^{\h\beta}\,.
\ee
From \eq{4.14} we can compute the tensor \eq{5.4} components,
\ba\n{5.6}
\Phi_{\h 0\h i}&=&{\cal B}_{\h i\h j;\h 0}x^{\h j}+\tfrac{1}{2}{\cal B}_{\h i\h j;\h k\h 0}x^{\h j}x^{\h k}\,,\non\\    
\Phi_{\h i\h j}&=&-\tfrac{1}{2}\left({\cal B}_{\h i\h k;\h j}-{\cal B}_{\h j\h k;\h i}\right)x^{\h k}\,.
\ea
Note that the covariant derivative of the gravitomagnetic field and its gradient with respect to $x^{\h 0}$ is just the derivative of the field \eq{4.15} and its gradient \eq{4.17} with respect to proper time $\tau$. 

Equation \eq{5.5} describes the gravitational spin-Hall effect of light in a local inertial frame in the space-time metric \eq{3.2}. Namely, the force term depends on the polarisation type via $\sigma$ value, that is the right- and left-polarised light components get deflected in the opposite transverse directions. It follows from the geometric optics condition $\varepsilon\ll1$ and the normal Fermi coordinates validity range \eq{3.3} that the components separation within the laboratory frame proper length $\ell$ is much less than $\varepsilon^2\lambdabar$. However, taking into account that handedness (helicity) of a circularly polarised light reverses under its normal reflection ($\sigma\to-\sigma$) and assuming that the tensor components \eq{5.6} do not change their sign in some time interval $\tau$, it is possible to achieve, by arranging a set of mirrors, that the light travels a sufficiently large distance by going back and forth in the laboratory frame, while its components are being under the influence of the transverse force of the same direction. This results in accumulation of their transverse separation and hence in a potential  observation of the gravitational spin-Hall effect.   

Let us now represent equation \eq{5.5} in terms of quantities measured at the given point on the light null geodesic. This can be achieved by decomposing the expression \eq{5.3} into parts parallel and orthogonal to the local $u^{\h\alpha}$ vector. The parallel part is derived by contracting \eq{5.3} with $u^{\h\alpha}$ twice and it is just an identity. The perpendicular part has two parts. The vector part is derived by contracting \eq{5.3} with $u^{\h\alpha}$ and then applying the projection operator \eq{4.5}. Contracting the vector component with the unit vector $n^{\h\alpha}$ and expressing the result in the local tetrad frame \eq{3.5} we derive
\be\n{5.7}
\omega_{,(0)}+n^{(a)}\omega_{,(a)}+w_{(a)}n^{(a)}+\theta_{(a)(b)}n^{(a)}n^{(b)}=\varepsilon\sigma n^{(a)}E_{(a)}\,.
\ee
Here 
\be\n{5.8}
\theta_{(a)(b)}=e^{\h\alpha}_{\,\,\, (a)}e^{\h\beta}_{\,\,\, (b)}\Pi_{\h\alpha}^{\h\mu}\Pi_{\h\beta}^{\h\nu}u_{(\h\mu;\h\nu)}
\ee
is expansion of the local congruence of time-like world lines $x^{\h i}=\text{const}$ and
\be\n{5.9}
E_{(a)}\equiv e^{\h\alpha}_{\,\,\, (a)}e^{\h\beta}_{\,\,\, (0)}\Phi_{\h\alpha\h\beta}
\ee
is the local ``electric" part of the tensor \eq{5.4}. The tensor part is derived by applying the projection operator \eq{4.5}  twice to the expression \eq{5.3} and contracting the derived expression with the unit vector $n^{\h\alpha}$. Expressing the result in the local tetrad frame \eq{3.5} we derive
\be\n{5.10}
\frac{D\BM{n}}{d\ell}=\frac{\nabla\omega}{\omega}-\frac{\BM{n}}{\omega}(\nabla\omega,\BM{n})-2[\BM{n}\times\BM{\Omega}]+\varepsilon\sigma[\BM{n}\times\BM{B}]\,.
\ee
Here 
\be\n{5.11}
\frac{Dn_{(a)}}{d\ell}=e^{\h\alpha}_{\,\,\, (a)}n^{\h\beta}n_{\h\beta;\h\alpha}\,,
\ee
where $d\ell$ is the proper distance element in the local space $\Sigma_{u}$,
\be\n{5.12}
B_{(a)}\equiv\tfrac{1}{2}e_{(a)(b)(c)}\Phi^{(b)(c)}\hhh \Phi_{(a)(b)}=e^{\h\alpha}_{\,\,\, (a)}e^{\h\beta}_{\,\,\, (b)}\Phi_{\h\alpha\h\beta}
\ee
is the local ``magnetic" part of the tensor \eq{5.4}, and $e_{(a)(b)(c)}$ is the Levi-Civita (pseudo) tensor normalised as $e_{(1)(2)(3)}=+1$. The scalar $(\BM{a},\BM{b})$ and vector $[\BM{a}\times\BM{b}]$ products between 3-vectors $\BM{a}$ and $\BM{b}$ lying in $\Sigma_{u}$ are defined in the usual way.

Equations \eq{5.7} and \eq{5.10} represent the null vector $\tilde{k}^{\h\alpha}$ evolution in terms of the local frequency $\omega$ and the unit space-like vector $\BM{n}$. Equation \eq{5.10} is analogous to the equation for light propagation in optically transparent medium with the refraction coefficient $n(\omega)=\omega$ rotating with the angular velocity $\BM{\Omega}$ in the presence of an external magnetic field, whose $curl$ is defined by $\BM{B}$. Indeed, from the gravitoelectromagnetic point of view (see, e.g. \cite{Mashhoon:1999nr}), the $g_{\h 0\h i}$ metric component in \eq{3.2} corresponds to a gravitoelectromagnetic vector potential, then the vorticity \eq{4.13} corresponds to a gravitomagnetic field, and finally, $\BM{B}$ corresponds to $curl$ of the gravitomagnetic field.  

\section{A role of observer}

Results of the previous sections show that a gravitomagnetic field plays a crucial role in the gravitational Faraday and spin-Hall effects of light. This field, as it follows from the expressions \eq{3.4}, \eq{4.9}, and \eq{4.14} depends on the observer's frame. This implies that these effects are observer-dependent phenomena. In this section, our goal is to illustrate a role of the local observer in the gravitational Faraday and spin-Hall effects of light.

As an illustrative example, we consider an inertial observer moving in the equatorial plane of a Kerr black hole. An orthonormal tetrad parallel-transported along a time-like geodesic in the Kerr geometry was constructed in \cite{marck1983solution}. By calculating the Weyl tensor of the Kerr space-time in the observer's tetrad frame we derive the gravitomagnetic field \eq{4.15},
\be\n{6.1}
{\cal B}_{\h\alpha\h\beta}=2B\cos\Psi\delta^{\h 1}_{(\h\alpha}\delta^{\h 2}_{\h\beta)}+2B\sin\Psi\delta^{\h 2}_{(\h\alpha}\delta^{\h 3}_{\h\beta)}\,,
\ee
where 
\be\n{6.2}
B=\frac{3M}{r^5}|Q|\sqrt{r^2+Q^2}\hh\dot\Psi=\pm\frac{EQ-a}{r^2+Q^2}\hh Q=aE-L\,.
\ee
Here $M$ is the black hole mass, $a$ is its reduced angular momentum, $E$ is the observer's energy per unit mass, $L$ is its azimuthal angular momentum, $\Psi$ is the angle of the tetrad rotation with respect to the Boyer-Lindquist coordinate basis, $\dot\Psi$ is its derivative with respect to the proper time, `$+$' stands for $Q>0$, and `$-$' for $Q<0$.
Gradient components \eq{4.17} of the gravitomagnetic field are the following:
\ba\n{6.3}
{\cal B}_{\h 1\h 1;\h 2}&=&-\frac{3M}{r^7}(r^3\dot{r}\sin(2\Psi)\pm X\cos^2\Psi\mp ar^2)\,,\non\\
{\cal B}_{\h 1\h 2;\h 1}&=&\frac{3M}{r^7}(r\dot{r}Z\sin\Psi\cos\Psi\mp Y\cos^2\Psi\pm r^2EQ)\,,\non\\
{\cal B}_{\h 1\h 2;\h 3}&=&-\frac{3M}{r^7}\cos\Psi(r\dot{r}Z\cos\Psi\pm Y\sin\Psi)\,,\non\\
{\cal B}_{\h 1\h 3;\h 2}&=&\frac{3M}{r^7}(r^3\dot{r}\cos(2\Psi)\mp X\sin\Psi\cos\Psi)\,,\non\\
{\cal B}_{\h 2\h 2;\h 2}&=&\pm\frac{3M}{r^7}(r^2(2EQ+a)+5aQ^2)\,,\non\\
{\cal B}_{\h 2\h 3;\h 1}&=&\frac{3M}{r^7}\sin\Psi(r\dot{r}Z\sin\Psi\mp Y\cos\Psi)\,,\non\\
{\cal B}_{\h 2\h 3;\h 3}&=&-\frac{3M}{r^7}(r\dot{r}Z\sin\Psi\cos\Psi\pm Y\sin^2\Psi\mp r^2EQ)\,,\non\\
{\cal B}_{\h 3\h 3;\h 2}&=&\frac{3M}{r^7}(r^3\dot{r}\sin(2\Psi)\mp X\sin^2\Psi\pm ar^2)\,,
\ea
where
\ba\n{6.4}
\dot{r}&=&\pm\frac{1}{r^2}\sqrt{(Er^2+aQ)^2-\Delta(r^2+Q^2)}\,,\non\\
\Delta&=&r^2+a^2-2Mr\hhh X=r^2(2EQ+3a)+5aQ^2\,,\non\\
Y&=&r^2(4EQ+a)+5aQ^2\hhh Z=r^2+5Q^2\,.
\ea
Thus, according to the expressions \eq{4.18} and \eq{5.5}, \eq{5.6}, the gravitational Faraday and spin-Hall effects can be detected in the observer's frame, except for light propagating in the black hole equatorial plane $(x^{\h 2}=0)$. 

The expressions above imply that both the black hole reduced angular momentum $a$ and the azimuthal angular momentum $L$ of the observer's frame contribute to these effects such that there is the symmetry
\be\n{6.5}
aE\to-L\hhh L\to-aE\,.
\ee
Thus, if the black hole were not rotating (a Schwarzschild black hole), then there would also be the gravitational Faraday effect due to the observer's angular momentum $L$, which couples to the light helicity via the curved due to the black hole mass $M$ space-time. Observers moving radially in a Schwarzschild space-time do not detect the gravitational Faraday rotation. On the other side, zero angular momentum observers $(L=0)$ of the Kerr space-time detect the gravitational Faraday rotation.

This situation is analogical to an electromagnetic field. Consider a charged particle placed at the origin of some frame. There is only an electrostatic field observed in that frame. However, in a frame moving with respect to the charged particle with non-zero angular momentum, both electric and magnetic fields are observed. The magnetic field vanishes if the frame moves radially with respect to the particle, i.e. if it has zero angular momentum. 

Let us now analyse this situation in case of a gravitoelectromagnetic field. The gravitoelectric field is defined by \cite{PhysRevD.33.915,Membrane,GRSolutions},  
\be\n{6.6}
{\cal E}_{\h\alpha\h\beta}\equiv C_{\h\alpha\h\gamma\h\beta\h\delta}u^{\h\gamma}u^{\h\delta}\,,
\ee
such that
\be\n{6.7}
{\cal E}_{\h\alpha\h\beta}={\cal E}_{\h i\h j}\delta^{\h i}_{\h\alpha}\delta^{\h j}_{\h\beta}\hhh {\cal E}_{\h i\h j}={\cal E}_{\h j\h i}\hhh {\cal E}^{\h i}_{\,\,\,\h i}=0\,,
\ee
and the gravitomagnetic field was already defined above [cf. the expressions \eq{4.14}, \eq{4.15}]. Consider a freely falling observer's (FFO) orthonormal frame in a Schwarzschild black hole space-time,
\ba\n{6.8}
e^{\alpha}_{\,\,\,\h0'}&=&\left[\frac{E}{f},\dot{r},0,\frac{L}{r^2}\right]\,,\non\\
e^{\alpha}_{\,\,\,\h1'}&=&\left[\frac{\dot{r}}{f},f^{1/2}+\frac{\dot{r}^2}{E+f^{1/2}},0,\frac{\dot{r}L}{r^2(E+f^{1/2})}\right]\,,\non\\
e^{\alpha}_{\,\,\,\h2'}&=&\left[0,0,\frac{1}{r},0\right]\,,\\
e^{\alpha}_{\,\,\,\h3'}&=&\left[\frac{L}{rf^{1/2}},\frac{\dot{r}f^{1/2}L}{r(E+f^{1/2})},0,\frac{1}{r}+\frac{L^2f^{1/2}}{r^3(E+f^{1/2})}\right]\,,\non
\ea
where 
\be\n{6.9}
\dot{r}=\pm\sqrt{E^2-f(1+L^2/r^2)}\hhh f=1-2M/r\,, 
\ee
$E$ and $L\geq0$ are the frame energy and azimuthal angular momentum, and the frame components are given in the standard Schwarzschild coordinates $(t,r,\theta,\phi)$. Consider also a fiducial observer's (FIDO) orthonormal frame, which is at rest with respect to the Schwarzschild black hole,
\ba\n{6.10}
e^{\alpha}_{\,\,\,\h0}&=&[f^{-1/2},0,0,0]\hhh e^{\alpha}_{\,\,\,\h1}=[0,f^{1/2},0,0]\,,\non\\
e^{\alpha}_{\,\,\,\h2}&=&[0,0,r^{-1},0]\hhh e^{\alpha}_{\,\,\,\h3}=[0,0,0,r^{-1}]\,.
\ea
Both frames are in the black hole equatorial plane $(\theta=\pi/2)$. Let the frame origins be located at the same point in space. Then frames of FFO and FIDO are related via Lorentz transformation \cite{MTW}:
\ba\n{6.11}
\Lambda_{\h0'}^{\,\,\,\h0}&=&\gamma=\frac{1}{\sqrt{1-v^2}}\hhh \Lambda_{\h i'}^{\,\,\,\h0}=\Lambda_{\h0'}^{\,\,\,\h i}=\gamma n^{\h i}v\,,\\
\Lambda_{\h i'}^{\,\,\,\h j}&=&(\gamma -1)n^{\h i}n^{\h j}+\delta_{\h i\h j}\hhh (n^{\h1})^2+(n^{\h2})^2+(n^{\h3})^2=1\non\,,
\ea
where parameters of the transformation are the following:
\ba\n{6.12}
n^{\h1}&=&\frac{\dot{r}}{Ev}\hhh n^{\h2}=0\hhh n^{\h3}=\frac{Lf^{1/2}}{Evr}\,,\non\\
v&=&\sqrt{1-f/E^2}\hhh \gamma=E/f^{1/2}\,.
\ea
Gravitoelectromagnetic field measured by FIDO is
\be\n{6.13}
{\cal E}_{\h i\h j}=\text{diag}\left(-\frac{2M}{r^3},\frac{M}{r^3},\frac{M}{r^3}\right)\hhh {\cal B}_{\h i\h j}=0\,.
\ee
To find how it is related to a gravitoelectromagnetic field in FFO frame we decompose it into scalar, vector, and tensor parts as follows:
\ba\n{6.14}
&&{\cal E}_{\h i\h j}={\cal E}^{\perp}_{\h i\h j}+{\cal E}^{\perp}_{n\h i}n_{\h j}+{\cal E}^{\perp}_{n\h j}n_{\h i}-\tfrac{1}{2}(\delta_{\h i\h j}-3n_{\h i}n_{\h j}){\cal E}_{nn}\,,\non\\
&&{\cal B}_{\h i\h j}={\cal B}^{\perp}_{\h i\h j}+{\cal B}^{\perp}_{n\h i}n_{\h j}+{\cal B}^{\perp}_{n\h j}n_{\h i}-\tfrac{1}{2}(\delta_{\h i\h j}-3n_{\h i}n_{\h j}){\cal B}_{nn}\,,
\ea
where the scalar normal-normal parts are
\be\n{6.15}
{\cal E}_{nn}={\cal E}_{\h i\h j}n^{\h i}n^{\h j}\hhh {\cal B}_{nn}={\cal B}_{\h i\h j}n^{\h i}n^{\h j}\,,
\ee
the vector normal-transverse parts are
\be\n{6.16}
{\cal E}^{\perp}_{n\h i}=\pi_{\h i}^{\,\,\h j}{\cal E}_{\h j\h k}n^{\h k}\hhh {\cal B}^{\perp}_{n\h i}=\pi_{\h i}^{\,\,\h j}{\cal B}_{\h j\h k}n^{\h k}\,,
\ee
and the tensor transverse-traceless parts are
\be\n{6.17}
{\cal E}^{\perp}_{\h i\h j}=\pi_{\h i}^{\,\,\h k}\pi_{\h j}^{\,\,\h l}{\cal E}_{\h k\h l}+\tfrac{1}{2}\pi_{\h i\h j}{\cal E}_{nn}\hh {\cal B}^{\perp}_{\h i\h j}=\pi_{\h i}^{\,\,\h k}\pi_{\h j}^{\,\,\h l}{\cal B}_{\h k\h l}+\tfrac{1}{2}\pi_{\h i\h j}{\cal B}_{nn}\,.
\ee
Here 
\be\n{6.18}
\pi_{\h i}^{\,\,\h j}=\delta_{\h i}^{\,\,\h j}-n_{\h i}n^{\h j}
\ee
is a projector on a two-dimensional plane perpendicular to the velocity vector $v^{\h i}=vn^{\h i}$. The same relations can be written in FFO frame. Then, using the definitions of the gravitoelectric and gravitomagnetic fields we derive relations between them in these frames. The normal-normal parts,
\be\n{6.19}
{\cal E}'_{nn}={\cal E}_{nn}\hhh {\cal B}'_{nn}={\cal B}_{nn}\,,
\ee
the vector normal-transverse parts,
\ba\n{6.20}
{\cal E}^{\perp}_{n\h i'}&=&\gamma\left({\cal E}^{\perp}_{n\h i}+e_{\h i\h j}^{\,\,\,\,\,\h k}v^{\h j}{\cal B}^{\perp}_{n\h k}\right)\,,\non\\
{\cal B}^{\perp}_{n\h i'}&=&\gamma\left({\cal B}^{\perp}_{n\h i}-e_{\h i\h j}^{\,\,\,\,\,\h k}v^{\h j}{\cal E}^{\perp}_{n\h k}\right)\,,
\ea
and the tensor transverse-traceless parts,
\ba\n{6.21}
{\cal E}^{\perp}_{\h i'\h j'}&=&\gamma^2\left[(1+v^2){\cal E}^{\perp}_{\h i\h j}+e_{\h i\h k}^{\,\,\,\,\,\,\h l}v^{\h k}{\cal B}^{\perp}_{\h l\h j}+e_{\h j\h k}^{\,\,\,\,\,\,\h l}v^{\h k}{\cal B}^{\perp}_{\h l\h i}\right]\,,\non\\
{\cal B}^{\perp}_{\h i'\h j'}&=&\gamma^2\left[(1+v^2){\cal B}^{\perp}_{\h i\h j}-e_{\h i\h k}^{\,\,\,\,\,\,\h l}v^{\h k}{\cal E}^{\perp}_{\h l\h j}-e_{\h j\h k}^{\,\,\,\,\,\,\h l}v^{\h k}{\cal E}^{\perp}_{\h l\h i}\right]\,.
\ea
Then, using the field \eq{6.13} in FIDO frame we derive the gravitomagnetic field in FFO frame,
\be\n{6.22}
{\cal B}_{\h1'\h2'}=\frac{3ML}{r^4}+\frac{3ML^2|L|f^{1/2}}{r^6(E+f^{1/2})}\hh {\cal B}_{\h2'\h3'}=-\frac{3ML^2\dot{r}}{r^5(E+f^{1/2})}\,.
\ee
The FIDO gravitoelectric field can be calculated in a similar way, however, it is not of our interest here. Note that in the weak gravitational field approximation the leading term of the gravitomagnetic field \eq{6.1} with $L=0$ coincides with the leading term of the gravitomagnetic field \eq{6.22} with $L=aE$. This is an illustration of the gravitational Larmor theorem \cite{Mashhoon:1999nr}. We see that FIDO of non-zero angular momentum $L$ detects the gravitomagnetic field and can observe the gravitational Faraday and spin-Hall effects in the Schwarzschild space-time. Thus, these effects are observer-dependent phenomena. 

\section{Discussion}

In this work we studied the gravitational Faraday and spin-Hall effects of light in a local inertial frame. Our main results are the expressions \eq{4.12}, \eq{4.18}, \eq{5.7}, and \eq{5.10}. These expressions imply that the effects are due to the gravitomagnetic field \eq{4.15} and its gradient \eq{4.17} measured in the local frame. Then we have shown that the gravitomagnetic field and thus the effects are frame dependent. In particular, one can consider an inertial zero angular momentum frame in a Kerr space-time where the effects can be observed, or an inertial frame with non-zero angular momentum in a Schwarzschild space-time, where the effects can be observed as well. This implies that in order to address the gravitational Faraday and spin-Hall effects of light to a gravitational field alone, one has to consider zero angular momentum observers. 

The advantage of the local approach is that one can study the gravitational Faraday and spin-Hall effects of light in a local inertial frame in any space-time region, whereas the global time-like vector field of observers defined through the Killing vector field in a stationary space-time exists only in the regions where the Killing vector field is time-like, e.g. outside a Kerr black hole ergosphere. We also recall that in order to use the presented here approach in arbitrary space-time, one has to find a local inertial frame \eq{3.2}, which requires finding a time-like geodesic and constructing an orthonormal tetrad parallel-transported along it. Then, the next problem is to find a null geodesics in that frame within the corresponding order of approximation. We shall leave these important issues outside the scope of this paper.

\acknowledgements

In Sec.~VI calculations of the expressions \eq{6.1}-\eq{6.5} were done by using GRTensor III computer algebra package.

\bibliography{biblio.bib}

\end{document}